\documentclass[12pt]{article}

\usepackage{graphics} 
\usepackage{amsmath}  
\usepackage{pdfsync}

\begin{document}

\textheight=24cm
\textwidth=16.5cm
\topmargin=-1.5cm
\oddsidemargin=-0.25cm.


\newcommand{\1}{{\'\i}}
\def\braket#1#2{\langle {#1} \mid {#2} \rangle}
\def\mn{\mu\nu}
\def\square{\kern1pt\vbox{\hrule height 1.2pt\hbox{\vrule width 1.2pt
   \hskip 3pt\vbox{\vskip 6pt}\hskip 3pt\vrule width 0.6pt}
   \hrule height 0.6pt}\kern1pt}
      \def\boxop{{\raise-.25ex\hbox{\square}}}

\renewcommand{\thefootnote}{\fnsymbol{footnote}}
\vskip .4cm
\begin{center}
{\Large\bf Photon-Graviton Amplitudes from}\\
{\Large\bf  the Effective Action}

\bigskip

{F. Bastianelli$^{a}$, O. Corradini$^{b}$, J. M. D\'avila$^{c}$ and 
C. Schubert$^{c}$}
\begin{itemize}
\item [$^a$]
{\it
Dipartimento di Fisica, Universit\`a di Bologna
  and\\ INFN, Sezione di Bologna, Via Irnerio 46, I-40126
  Bologna, Italy
  }
\item [$^b$] 
 {\it 
Centro de Estudios en F\'isica y Matem\'aticas B\'asicas y Aplicadas\\
Universidad Aut\'onoma de Chiapas\\ C.P. 29000, Tuxtla Guti\'errez, M\'exico
}
\item [$^c$]
{\it 
Instituto de F\'{\i}sica y Matem\'aticas
\\
Universidad Michoacana de San Nicol\'as de Hidalgo\\
Edificio C-3, Apdo. Postal 2-82\\
C.P. 58040, Morelia, Michoac\'an, M\'exico\\
}
\end{itemize}
\end{center}

\begin{center}
{Talk given by C. Schubert at {\it Supersymmetries and Quantum Symmetries - SQS`2011}, JINR Dubna, July 18 - 23, 2011
(to appear in the Proceedings)}
\end{center}

\noindent
{\bf Abstract:}
We report on the status of an ongoing effort to calculate the complete one-loop low-energy effective actions in 
Einstein-Maxwell theory with a massive scalar or spinor loop, and to use them for obtaining the explicit form of 
the corresponding M-graviton/N-photon amplitudes. 
We present explicit results for the effective actions at the one-graviton four-photon level, and for the
amplitudes at the one-graviton two-photon level. As expected on general grounds, these amplitudes 
relate in a simple way to the corresponding four-photon amplitudes. 
We also derive the gravitational Ward identity for the 1PI one-graviton -- N photon amplitude. 

%


\renewcommand{\thefootnote}{\protect\arabic{footnote}}

\setcounter{page}{1}
\setcounter{footnote}{0}

\section{Introduction: }
\renewcommand{\theequation}{1.\arabic{equation}}
\setcounter{equation}{0}

In string theory, the prototypical example of relations between gravity and gauge theory
amplitudes are the ``KLT'' relations discovered by Kawai et al. \cite{kalety}. Schematically,
they are of the form 

$$({\rm gravity\, amplitude}) \sim  ({\rm gauge\, amplitude})^2 $$ 

\noindent
and they follow naturally 
from the factorization of the graviton vertex operator into a product of two
gauge boson vertex operators (see, e.g., \cite{bernrev})

$$V^{\rm closed} = V^{\rm open}_{\rm left}\bar V^{\rm open}_{\rm right}$$

These string relations induce also relations in field theory. 
For example, at four and five point one has \cite{bernrev}

\vspace{-10pt}

\begin{eqnarray}
M_4(1,2,3,4) &=& - i s_{12}A_4(1,2,3,4)A_4(1,2,4,3)\nonumber\\
M_5(1,2,3,4,5) &=& is_{12}s_{34}A_5(1,2,3,4,5)A_5(2,1,4,3,5)\nonumber\\
&&+ is_{13}s_{24}A_5(1,3,2,4,5)A_5(3,1,4,2,5)\nonumber\\
\label{45point}
\end{eqnarray}
Here the $M_n$ are $n$ - point tree-level graviton amplitudes, and the
$A_n$ are  (colour-stripped) tree-level gauge theory amplitudes. The 
$s_{ij} = (k_i+k_j)^2$ are kinematical invariants.

Although the work of \cite{kalety} was at the tree level, by unitarity
those tree level relations induce also identities at the loop level. 
By now, many relations between graviton and gauge amplitudes have been
derived along these lines at the one loop level and beyond;
see \cite{bcdjr} and refs. therein.
Presently a key issue here is the possibility that 
the finiteness of $N=4$ SYM theory may extend to  $N=8$ Supergravity
(see \cite{bcdjr,fermar} and P. Vanhove's talk at this conference). Finiteness of a quantum field theory usually
implies extensive cancellations between Feynman diagrams, and it is presently
still not well-understood what are the precise extent and origin of such cancellations in the Supergravity case.

In this respect, gravity amplitudes are more similar to QED amplitudes than to
nonabelian amplitudes, since colour factors greatly reduce the
potential for cancellations between diagrams. 
In QED, there are many cases of surprising cancellations between 
diagrams. A famous case is 
the three-loop QED $\beta$ - function coefficient, involving the sum of diagrams shown in fig. 1.

\begin{center}
\begin{figure}[ht]
\includegraphics{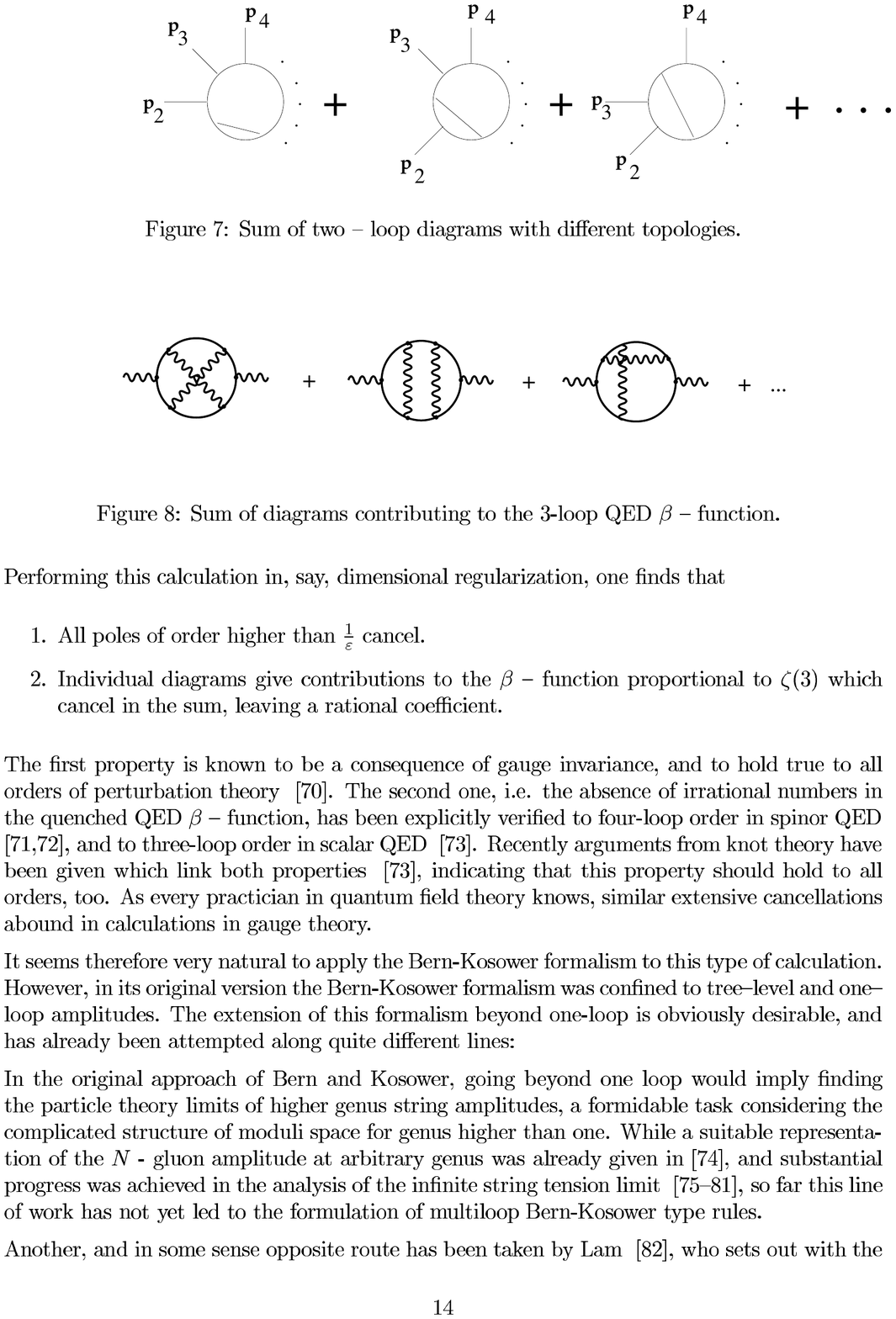}
\caption{{Sum of diagrams for  the three-loop QED photon propagator}}
\label{rosner}
\end{figure}
\end{center}

\vspace{-20pt}

\noindent
As discovered by Rosner in 1967 \cite{rosner}, individual diagrams give contributions to the $\beta$ - function
coefficient that involve $\zeta(3)$, however those terms cancel out, leaving a simple rational number
for the sum of diagrams. 
Such cancellations are usually attributed to gauge invariance, since they generally appear inside gauge invariant sets of graphs. 
Even for QED, little is still known about the influence of these cancellations
on the large-order behaviour of the QED perturbation series
\cite{cvitanovic1977,60}. 
For recent gravity-inspired studies of the structure of QED amplitudes see \cite{babiva,brtrvi,badhen}.

Considering the enormous amount of work that has been done on the structural relationships between gauge and
gravity amplitudes, it is surprising that relatively few results exist for mixed graviton-gluon or
graviton-photon amplitudes \cite{bedewo,holstein2006,arcode}.
In this talk, we report on the status of an ongoing effort to calculate the complete one-loop low-energy effective actions in 
Einstein-Maxwell theory with a massive scalar or spinor loop, and to use them for obtaining the explicit form of 
the corresponding M-graviton/N-photon amplitudes \cite{76,79,bcds}. The talk is organized as follows:
In chapter 2 we will shortly summarize what is presently known
about the QED $N$ photon amplitudes.
In chapter 3 we summarize the results of \cite{76,79} 
on the one-loop effective action in Einstein-Maxwell theory, 
and also improve somewhat on the form of its
one-graviton four-photon part as compared to \cite{79}. 
Chapter 4 is devoted to the graviton - photon - photon amplitude.
Our conclusions are presented in chapter 5.

\section{Properties of the QED $N$ photon amplitudes}
\label{propNphoton}
\renewcommand{\theequation}{2.\arabic{equation}}
\setcounter{equation}{0}

We shortly summarize what is known about the $N$ photon-amplitudes in scalar and spinor QED
(results given refer to the spinor case unless stated otherwise). 

Although the one-loop  four-photon amplitude 
was calculated by Karplus and Neumann already in 1950 \cite{karneu}, progress towards higher leg or
multiloop photon amplitudes has been extremely slow. 
The one-loop six-photon amplitude (recall that by Furry's theorem there
are no amplitudes with an odd number of photons) was obtained only quite recently \cite{bhgm},
and only for the massless case. On-shell amplitudes for gauge bosons are nowadays generally given in the
helicity eigenstate decomposition; using CP invariance, the six-photon amplitude then has four independent components,
which can be chosen as 
$A(++++++),A(+++++-),A(++++--),A(+++---)$
(in the gluonic case there will be more independent
components since the ordering of the legs matters).

Apart from these explicit low order calculations, there are also a number of all-$N$ results. First, for massless QED there is
Mahlon's vanishing theorem \cite{mahlon}, stating that  
$A_N(+++ \ldots ++) = A_N(-++\cdots ++) = 0$ for $N>4$. Mahlon also obtained a closed formula for 
the first non-vanishing case of two negative helicities $A(--+\ldots +)$ in terms of dilogarithms \cite{mahlon--}.

More recently, Badger et al. \cite{babiva} have shown that the massless $N$ photon amplitudes
for $N\geq 8$ fulfill the ``no triangle"' property, that is, after the usual reduction from tensor to scalar
integrals  they involve only box integrals but not triangle ones. 
This is analogous to the ``no triangle'' property of ${\cal N}=8$ supergravity \cite{bievan}, 
which is important for the possible finiteness of that theory.

An explicit all - $N$ calculation is possible for the low-energy limit of the massive photon amplitudes, where
all photon energies are small compared to the electron mass,  $\omega_i \ll m $.
The information on the $N$ photon amplitudes in this limit is contained in the well-known Euler-Heisenberg \cite{eulhei}
(for spinor QED) resp. Weisskopf \cite{weisskopf} (for scalar QED) effective Lagrangians:

 \begin{eqnarray}
{\cal L}_{\rm spin}(F) &=& - \frac{1}{8\pi^2}
\int_0^{\infty}{dT\over T^3}
\,{\rm e}^{-m^2T} 
\biggl[
{(eaT)(ebT)\over {\rm tanh} (eaT){\rm tan} (ebT)} 
- {1\over 3}(a^2-b^2)T^2 -1
\biggr]
\nonumber\\
{\cal L}_{\rm scal}(F)&=&  {1\over 16\pi^2}
\int_0^{\infty}{dT\over T^3}
\,{\rm e}^{-m^2T} 
\biggl[
{(eaT)(ebT)\over \sinh(eaT)\sin(ebT)} 
+{1\over 6}(a^2-b^2)T^2 -1
\biggr]\nonumber\\
\label{eulhei}
\end{eqnarray}
Here $T$ is the proper-time of the loop scalar or spinor particle and $a,b$  are defined by 
$a^2-b^2 = B^2-E^2,\quad  ab = {\bf E}\cdot {\bf B}$. Extracting the 
on-shell amplitudes from the effective action is a standard procedure in quantum field
theory. In the helicity decomposition,   
one finds \cite{56}

\begin{eqnarray}
A_{\rm spin}^{(EH)}
[\varepsilon_1^+;\ldots ;\varepsilon_K^+;\varepsilon_{K+1}^-;\ldots ;\varepsilon_N^-]
&=&
-{m^4\over 8\pi^2}
\Bigl({2ie\over m^2}\Bigr)^N(N-3)!
\nonumber\\&&\hspace{-130pt}\times
\sum_{k=0}^K\sum_{l=0}^{N-K}
(-1)^{N-K-l}
{B_{k+l}B_{N-k-l}
\over
k!l!(K-k)!(N-K-l)!}
\chi_K^+\chi_{N-K}^- 
\nonumber\\
\label{Aspin}
\end{eqnarray}
and a similar formula for the scalar QED case \cite{56}.
Here the $B_k$ are Bernoulli numbers, and the variables $\chi_K^{\pm}$ are written, in spinor helicity
notation (our spinor helicity conventions follow \cite{dixonrev})

\begin{eqnarray}
\chi_K^+ &=&
{({\frac{K}{2}})!
\over 2^{K\over 2}}
\Bigl\lbrace
[12]^2[34]^2\cdots [(K-1)K]^2 + {\rm \,\, all \,\, permutations}
\Bigr\rbrace
\nonumber\\
\chi_K^- &=&
{({\frac{K}{2}})!
\over 2^{K\over 2}}
\Bigl\lbrace
\langle 12\rangle ^2\langle 34\rangle ^2\cdots \langle(K-1)K\rangle^2 + {\rm \,\, all \,\, permutations}
\Bigr\rbrace\nonumber\\
 \label{defchi}
\end{eqnarray} 
These variables appear naturally in the low energy limit. 
Since they require even numbers of positive and negative helicity
polarizations, in this low energy limit we find a  ``double Furry theorem'':
Only those helicity components are non-zero where both the number of
positive and negative helicity photons are even. It is easy to show
that this even holds true to all loop orders. 
For the MHV (``maximally helicity violating'' = ``all $+$'' or ``all $-$'') case (\ref{Aspin})
and its scalar analogue
imply that the scalar and spinor amplitudes differ only by the global factor of $-2$ for
statistics and degrees of freedom:

\begin{eqnarray}
A_{\rm spin}^{(EH)}
[\varepsilon_1^+;\ldots ;\varepsilon_N^+]
= -2
A_{\rm scal}^{(EH)}
[\varepsilon_1^+;\ldots ;\varepsilon_N^+]
\label{relscalspin}
\end{eqnarray}
This well-known relation is actually true also away from the low-energy limit, 
and can be explained by the fact that the MHV amplitudes correspond to a 
self-dual background, in which the Dirac operator has
a quantum-mechanical supersymmetry \cite{dufish}.

\section{One-loop effective action in Einstein-Maxwell theory}
\label{EMEA}
\renewcommand{\theequation}{3.\arabic{equation}}
\setcounter{equation}{0}

The calculation of the one-loop effective action in Einstein-Maxwell theory
is usually done using heat kernel techniques. The first calculation of relevance
in our present context of the on-shell photon-graviton amplitudes  
was performed by Drummond and Hathrell \cite{druhat}, who obtained 
the terms in the fermionic effective Lagrangian involving one curvature tensor and
two field strength tensors:

\begin{eqnarray}
{\cal L}_{\rm spin}^{(DH)} &=& 
\frac{1}{180 (4\pi)^2m^2} \bigg(
5 R F_{\mu\nu}^2 
-26 R_{\mu\nu} F^{\mu\alpha} F^\nu{}_\alpha
+2  R_{\mu\nu\alpha\beta}F^{\mu\nu}F^{\alpha\beta} 
+24 (\nabla^\alpha F_{\alpha\mu})^2  
\bigg )
\nonumber\\
\label{drumhath}
\end{eqnarray}
(here and in the following we will often absorb the electric charge $e$ into the field strength tensor
$F$).

Recently, some of the present authors used the worldline formalism \cite{41,61} to obtain
effective Lagrangians that contain the full information on the low energy limit of the one-loop 
one graviton - $N$ photon amplitudes in Einstein-Maxwell theory with a scalar or spinor loop \cite{76}.
Those Lagrangians, which generalize the above Euler-Heisenberg and Weisskopf Lagrangians
as well as the Drummond-Hathrell Lagrangian (\ref{drumhath}), were obtained in terms of
two-parameter integrals of trigonometric power series in the field strength
matrix; to extract from them the part relevant at the one-graviton $N$ photon level,
one has to expand the integrand in powers of $F_{\mu\nu}$ up to $F^n$, after which the integrals
are polynomial and thus can be done by computer. After this, the gauge and gravitational Bianchi
identities can be used to greatly reduce the number of terms. 
At the one graviton -- two photon level one finds \cite{76}

\begin{eqnarray}
{\cal L}_{\rm scal}^{h\gamma\gamma}&=&
{1\over 360\, m^2 (4\pi)^2}
\Bigl\lbrack
5(6\xi -1)RF_{\mn}^2 +4R_{\mn}F^{\mu\alpha}F^{\nu}{}_{\alpha}
-6R_{\mu\nu\alpha\beta}F^{\mn}F^{\alpha\beta} 
\nonumber\\
&& \qquad\qquad\qquad
-2(\nabla^{\alpha}F_{\alpha\mu})^2
-8(\nabla_{\alpha}F_{\mn})^2
-12F_{\mn}\square F^{\mn}
\Bigr\rbrack\label{3pointscal}\\
{\cal L}_{\rm spin}^{h\gamma\gamma}
&=& {1\over 180\, m^2 (4\pi)^2}
\Bigl\lbrack
5RF_{\mn}^2 
-4R_{\mn}F^{\mu\alpha}F^{\nu}{}_{\alpha}
-9R_{\mu\nu\alpha\beta}F^{\mn}F^{\alpha\beta} 
\nonumber\\
&& \qquad\qquad\qquad
+2(\nabla^{\alpha}F_{\alpha\mu})^2
-7(\nabla_{\alpha}F_{\mn})^2
-18F_{\mn}\square F^{\mn}
\Bigr\rbrack 
\label{3pointspin}
\end{eqnarray}
The result for the spinor loop case differs from the Drummond-Hathrell Lagrangian by a total derivative term \cite{76}. 
The parameter $\xi$ appearing in the scalar case represents a non-minimal coupling to gravity.
At the next, $N=4$ level (there are no amplitudes with an odd number of photons by an extension of 
Furry's theorem to the photon-graviton case) this procedure is already quite laborious. It was carried through in
Ref. \cite{79}, but here we give the results in a slightly more compact form than was obtained there:

\begin{eqnarray}
{\cal L}_{\rm spin}^{h,4\gamma}
&=&-\frac{1}{8\, \pi^2}\frac{1}{m^6}\Bigg[-\frac{1}{432}R(F_{\mu \nu})^4 +\frac{7}{1080}R\, \mbox{tr}[F^4]
-\frac{1}{945 }R_{\alpha \beta}(F^4)^{\alpha \beta}\nonumber\\
&&-\frac{1}{540}R_{\alpha \beta}(F^2)^{\alpha \beta}(F_{\gamma \delta})^2+\frac{4}{135}R_{\alpha \mu \beta \nu}(F^3)^{\alpha \mu}F^{\beta \nu}
+\frac{1}{108}R_{\alpha \mu \beta \nu} F^{\alpha \mu} F^{\beta \nu}(F_{\gamma \delta})^2 \nonumber\\
&&+\frac{7}{270}(F^3)^{\mu \nu}\,\square F_{\mu \nu}+\frac{1}{108}F^{\mu \nu}\square F_{\mu \nu}(F_{\gamma \delta})^2 +\frac{1}{270}F_{\mu \nu ; \alpha \beta}(F^2)^{\alpha \beta}F^{\mu \nu} \nonumber\\
&&-\frac{1}{540}(F_{\alpha \beta ; \gamma})^2 (F_{\mu \nu})^2
-\frac{1}{945}F_{\mu \nu;\alpha}\,F^{\mu \nu}{}{}_{;\beta}(F^2)^{\alpha \beta}
-\frac{11}{945}F_{\alpha \beta ; \gamma}F^{\ \beta ; \gamma}_{\mu}(F^2)^{\alpha \mu} \nonumber\\
&&-\frac{2}{189}F_{\alpha \beta ; \gamma}F_{\mu \nu ;}^{\ \ \ \gamma}F^{\alpha \mu}F^{\beta \nu}
-\frac{2}{189}F_{\alpha \beta ; \gamma}F_{\mu \ \ ; \delta}^{\ \alpha}F^{\beta \mu}F^{\gamma \delta} \Bigg]
\end{eqnarray}

\begin{eqnarray}\label{LscalF4new}
{\cal L}_{\rm scal}^{h,4\gamma}&=&\frac{1}{16\, \pi^2}\frac{1}{m^6}\Bigg[-\frac{1}{144}\left(\bar\xi +\frac{1}{12}\right)R(F_{\mu \nu})^4
-\frac{1}{180}\left(\bar\xi +\frac{1}{12}\right)R\, \mbox{tr}[F^4] 
-\frac{1}{945 }R_{\alpha \beta}(F^4)^{\alpha \beta}\nonumber\\
&&+\frac{1}{1080}R_{\alpha \beta}(F^2)^{\alpha \beta}(F_{\gamma \delta})^2
-\frac{1}{270}R_{\alpha \mu \beta \nu}(F^3)^{\alpha \mu}F^{\beta \nu}
+\frac{1}{432}R_{\alpha \mu \beta \nu} F^{\alpha \mu} F^{\beta \nu}(F_{\gamma \delta})^2 \nonumber\\
&&-\frac{1}{540}(F^3)^{\mu \nu}\,\square F_{\mu \nu}
+\frac{1}{432}F^{\mu \nu}\square F_{\mu \nu}(F_{\gamma \delta})^2 
-\frac{1}{540}F_{\mu \nu ; \alpha \beta}(F^2)^{\alpha \beta}F^{\mu \nu} \nonumber\\
&&+\frac{1}{1080}(F_{\alpha \beta ; \gamma})^2 (F_{\mu \nu})^2
-\frac{1}{945}F_{\mu \nu;\alpha}\,F^{\mu \nu}{}{}_{;\beta}(F^2)^{\alpha \beta} 
-\frac{1}{1890}F_{\alpha \beta ; \gamma}F^{\ \beta ; \gamma}_{\mu}(F^2)^{\alpha \mu} \nonumber\\
&&+\frac{1}{1890}F_{\alpha \beta ; \gamma}F_{\mu \nu ;}^{\ \ \ \gamma}F^{\alpha \mu}F^{\beta \nu}
+\frac{1}{1890}F_{\alpha \beta ; \gamma}F_{\mu \ \ ; \delta}^{\ \alpha}F^{\beta \mu}F^{\gamma \delta} \Bigg]\,,\,\,\,\,\,\,\,\,\,\,\,\,
\end{eqnarray}
($\bar\xi = \xi - \frac{1}{4}$).
This improvement over the formulas given in Ref. \cite{79} is due to the following consequence of the Bianchi
identities, that had been overlooked in the list of identities used there:

\begin{eqnarray}\label{new}
R_{\alpha \mu \beta \nu}\,(F^2)^{\alpha \beta}\,(F^2)^{\mu \nu}&=&-\frac{1}{2} F_{\mu \nu ;  \alpha \beta}\,(F^2)^{\alpha \beta}\,F^{\mu \nu}
-\frac{1}{2}R_{\alpha \mu \beta \nu}\,(F^3)^{\alpha \mu}\,F^{\beta \nu}
\nonumber\\
&& -F_{\mu \nu ;  \alpha \beta}\,(F^2)^{\alpha \nu}\,F^{\beta \mu}\,.\nonumber\\
\end{eqnarray}

\section{The graviton - photon - photon amplitude and its properties}
\label{phograv}
\renewcommand{\theequation}{4.\arabic{equation}}
\setcounter{equation}{0}

\begin{figure}[h]
  \centering
    \includegraphics{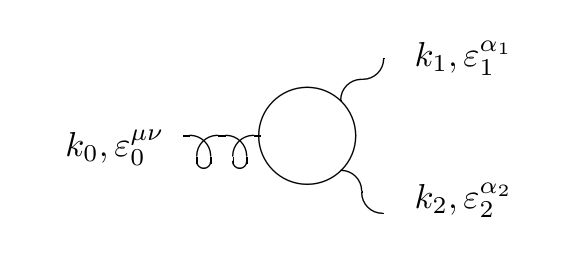}
      \caption{Graviton-photon-photon diagram}
      \label{gpp}
\end{figure}

We proceed to the simplest amplitude case, the graviton-photon-photon amplitude shown in fig. \ref{gpp}.

Getting its low energy limit from  the three-point Lagrangians (\ref{3pointscal}), (\ref{3pointspin}) (or equivalently from (\ref{drumhath})
in the spinor case) is straightforward.  In the helicity basis, and using the standard factorization of the graviton polarization tensor
in terms of vector polarizations,
$\varepsilon_{0\mu\nu}^{\pm\pm}(k_0) = \varepsilon_{\mu}^{\pm}(k_0)\varepsilon_{\nu}^{\pm}k_0)$,
one finds that only the ``all $+$'' and ``all $-$'' components are nonzero:

\begin{eqnarray}
A_{\rm spin}^{(++;++)}&=&\frac{\kappa\,e^2}{90 (4\pi)^2m^2}\,[01]^2\,[02]^2\nonumber\\ 
A_{\rm spin}^{(- - ; - - )}&=&\frac{\kappa\,e^2}{90 (4\pi)^2m^2}\,\langle 01 \rangle^2\,\langle 02 \rangle^2
\nonumber\\ 
\label{gppcomp}
\end{eqnarray}
Here the first upper index pair refers to the graviton polarization, and $\kappa$ is the gravitational coupling constant.
Moreover, those components fulfill the MHV relation (\ref{relscalspin}),

\begin{eqnarray}
A_{\rm spin}^{(++;++)}&=&(-2)\,A_{\rm scal}^{(++;++)}\nonumber\\
A_{\rm spin}^{(- - ; - - )}&=&(-2)\,A_{\rm scal}^{(- - ; - - )} \nonumber\\
\label{gppMHV}
\end{eqnarray}
Also, these graviton-photon-photon amplitudes relate to the (low energy) four photon amplitudes in the following way:
From (\ref{Aspin}), (\ref{defchi}) the only non-vanishing components of those are

\begin{eqnarray}
A^{++++}[k_1,k_2,k_3,k_4] &\sim& [12]^2[34]^2 +  [13]^2[24]^2 +  [14]^2[23]^2 \nonumber\\
A^{++--}[k_1,k_2,k_3,k_4] &\sim& [12]^2 \langle34\rangle^2 \nonumber\\
A^{----}[k_1,k_2,k_3,k_4] &\sim&
\langle12\rangle^2\langle34\rangle^2+\langle13\rangle^2\langle24\rangle^2+ \langle14\rangle^2\langle23\rangle^2 \nonumber\\
\label{4photcomp}
\end{eqnarray}
Replacing  $k_1 \to k_0, k_2 \to k_0$ in the 4 photon amplitudes, the middle one of these three components becomes zero,
and the remaining ones become proportional to the corresponding components of (\ref{gppcomp}),

\begin{eqnarray}
A^{++++}[k_0,k_0,k_3,k_4] \sim& 2[03]^2[04]^2 & \sim A^{++;++}[k_0,k_3,k_4] \nonumber\\
A^{----}[k_0,k_0,k_3,k_4] \sim& 2\langle03\rangle^2\langle04\rangle^2& \sim 
A^{--;--}[k_0,k_3,k_4] \nonumber\\
\label{gravtophot}
\end{eqnarray}
Thus effectively two photons have coalesced to form a graviton, clearly a result in the spirit of the KLT relations.

At the next level of one graviton and four photons, the conversion of the effective action into amplitudes becomes
already extremely laborious. Moreover, here there are already one-particle reducible contributions to the
amplitudes, with the graviton attached to a photon, and those are essential to arrive at a well-defined helicity decomposition. This is because
the 1PI amplitudes are transversal in the photon indices, but not in the graviton ones; rather, one has the inhomogeneous Ward identity \cite{bcds}

\begin{eqnarray}
 2 k_{0\mu} A^{\mu\nu,\alpha_1\ldots \alpha_N}[k_0,\ldots,k_N]
&=& - \sum_{i=1}^N A^{\mu\alpha_1\ldots \widehat{\alpha_i}\ldots \alpha_N}
[k_0+k_i,k_1,\ldots,\widehat{k_i},\ldots,k_N]
\nonumber\\&&\qquad\qquad\times
 (\delta^{\alpha_i}_\mu k_i^\nu -\eta^{\alpha_i\nu} k_{i\mu} )
 \label{wardgrav}
\end{eqnarray}
(where a `hat' means omission)
which connects the one graviton - N photon amplitudes to the N photon amplitudes.

\section{Conclusions}
\label{conclusions}
\renewcommand{\theequation}{5.\arabic{equation}}
\setcounter{equation}{0}

We have presented here the first results of a systematic study of the mixed one-loop photon - graviton amplitudes with
a scalar or spinor loop in the low energy limit. At the one graviton - two photon level, we find a KLT like factorization
of the graviton into two photons. If this type of factorization persisted for higher points, it would imply that, in the low energy limit, the full information
on the  $M$ graviton -- $N$  photon amplitudes is contained in the $N+2M$ photon amplitudes. However, the three-point result
may not be representative due to the absence of one-particle reducible contributions.  The situation will be clearer after the
completion of the one graviton - four photon calculation, which is presently in progress. 

\bigskip

\noindent
{\bf Acknowledgements:}  The work of F.B. was supported in part by the MIUR-PRIN contract 2009-KHZKRX.
 The work of O.C. was
partly funded by SEP-PROMEP/103.5/11/6653. C.S. was supported by
CONACYT grant CB 2008 101353.

\end{document}